

\let\miguu=\footnote
\def\footnote#1#2{{\parindent=0pt\baselineskip=14pt\miguu{#1}{#2}}}
\magnification=1200
\raggedbottom
\hsize=6 true in
\hoffset=0.27 true in
\vsize=8.5 true in
\voffset=0.28 true in
\baselineskip=14pt
\centerline {{\bf A SPECIMEN OF THEORY CONSTRUCTION}}
\centerline {{\bf FROM QUANTUM GRAVITY}\footnote{$\dagger$}
{\noindent
the text of an address delivered at the Thirteenth Annual Symposium in the
Philosophy of Science, entitled {\it How Theories are Constructed: The
Methodology of Scientific Creativity}, held at Greensboro, North Carolina,
March, 1989; to be published in {\it The Creation of Ideas in Physics},
ed. Jarrett Leplin (Kluwer Academic Publishers, Dordrecht, 1995)}}
\bigskip
\baselineskip=12pt
\centerline {\it Rafael D. Sorkin}
\centerline {\it Department of Physics}
\centerline {\it Syracuse University}
\centerline {\it Syracuse, NY 13244-1130}
%
\baselineskip=12pt
\bigskip
\centerline{\bf Abstract}
\medskip
\leftskip=1.5truecm\rightskip=1.5truecm     
\noindent
I describe the history of my attempts to arrive at a discrete substratum
underlying the spacetime manifold, culminating in the hypothesis that
the basic structure has the form of a partial-order (i.e. that it is  a
causal set).

\leftskip=0truecm\rightskip=0truecm         


\baselineskip=14pt
\bigskip

Like the other speakers in this session, I too am here much more as a
working scientist than as a philosopher.  Of course it is good to remember
Peter Bergmann's description of the physicist as ``in many respects a
philosopher in workingman's\footnote{*}
{\noindent The quotation comes from an earlier time. Today Peter would no
doubt use `working person', or some other non-sexist locution.}
clothes'', but today I'm not going to change into a white shirt and attempt
to draw philosophical lessons from the course of past work on quantum
gravity.  Instead I will merely try to recount a certain part of my own
experience with this problem, explaining how I arrived at the idea of what
I will call a causal set.  This and similar structures have been proposed
more than once as discrete replacements for spacetime.
My excuse for not telling you also how others arrived at essentially the
same idea [1] is naturally that my case is the only one I can hope to
reconstruct with even minimal accuracy.

\eject

\noindent{\bf The background of the problem}
\nobreak

Before describing the development I have just referred to, I should
probably tell you what a causal set is. Before doing that, however, let
me begin by saying a few words about the problem of quantum gravity
itself. What people somewhat misleadingly call by this name is really
the problem of restoring to physics the unified foundation it has lacked
since the beginning of this century.  If we adopt a slightly mythical
view of how science progresses, we can imagine that a new theory begins
to be constructed when too many experimental results accumulate in
conflict with the old theory.  A better theoretical understanding will
then emerge, but it may take some time to put  the pieces of this new
understanding together in a coherent manner. It may even happen that
these pieces cannot be mutually reconciled at all without some
fundamental extension of theory that would allow the
contradiction-among-the-parts to be dissolved within the context of a
more comprehensive whole.

The present situation of ``fundamental physics'' is similar to that I
have just described. Both Quantum Theory and General Relativity are
consistent with the facts they were created to explain, but they are
not consistent with each other. That this contradiction is purely
internal to theory has meant until very recently that only people with
a philosophical bent have taken the quantum gravity problem very
seriously. (It is thus a very appropriate topic to be discussed at a
philosophy of science conference.) Recently this neglect has given way
to intense interest; but it still cannot be said that we have any
direct conflict between experiment and accepted theory to guide us.

Why is it that quantum gravity suffers from such a lack of clearly
relevant experimental data, and what kind of experiment or observation
{\it could} be expected to provide such data?  Historically you could
say that Quantum Theory deals with the very small and General
Relativity with the very large, but the essence of the distinction is
not really one of size. Rather, ``the quantum of action'' is in general
important whenever no more than a few degrees of freedom are
excited,\footnote{*}{
\noindent This generally, but by no
means always, means that
only a few particles are involved.}
while gravity $-$ or in other words
General Relativity $-$ is important whenever a large enough amount of
energy is compressed into a small enough space. More specifically,
gravity is important when the ratio $Gm/rc^2$ is of order unity, where
$m$ is the total mass-energy, $r$ is the radius of the region into which it
has been compressed, and $G$ and $c$ are respectively the gravitational
constant and the speed of light. Actually we can sometimes notice
gravity in less extreme conditions than this, but to do so takes very
precise measurements, or a very long time, such as the time it takes a
satellite to circle the earth (which is indeed huge compared to the
radius of the earth in light units).

In any case, a typical object for which gravity always {\it will} be
important is a black hole. Now for this object, we can count the number
of its states $N$ using the known value of its entropy $S$ and the basic
formula (or definition if you will) $S = k  \log N$. The result is that
$N$ is gigantic for an astrophysical black hole, but of order $1$ when
the black hole's radius approaches the so-called Planck length of about
$10^{-32}$cm. If we could directly observe nature at this length-scale,
we would expect to see quantum black holes, and more generally to see
everything which occurs exhibiting both quantum and gravitational
features. However, since the smallest lengths to which we have so far
managed to penetrate by means of particle accelerators are around
$10^{-16}$cm., there is little hope of doing laboratory experiments in
quantum gravity for a long time to come.

The problem, then, is not that we make  wrong predictions about
processes which we haven't seen yet anyway,  but that we  fail to make
any predictions at all. The dynamical principles learned from quantum
mechanics just seem to be incompatible with the idea that gravity is
described by a metric field on a continuous manifold.  When we try to
combine these elements in a way similar to how we have ``quantized''
non-gravitational field theories, we run into apparently insurmountable
technical and conceptual problems, of which I will mention only three.

First the quantum amplitudes resulting from such a ``quantization'' turn
out to be ``non-renormalizable'', which means in effect that the theory
they define ceases to make sense at short  distances $-$ very likely
just at those distances where we expect to see quantum gravitational
effects in the first place!  Moreover the standard formulations of
quantum field theories rely  on the existence of a ``background'' notion
of time with respect to which dynamical evolution can be defined,
whereas Relativity makes time itself part of the dynamics.  This leads
both to difficulties in interpreting the formalism, and to technical
problems in setting up what is called the Hilbert space metric. Finally
the quantum Uncertainty Principle seems to combine with the General
Relativistic connection between mass and spacetime-curvature  in such a
way that any Gedanken-experiment attempting to measure the metric at
short distances gets trapped in a vicious circle: the more accuracy you
try for, the greater the uncontrollable disturbance you induce in the
geometry you are trying to measure.

The difficulties just mentioned arise when you attempt to unite Quantum
Field Theory with General Relativity, but actually each of these two
theories already has its own internal contradictions. Unquantized
gravity gives rise to singularities where the Einstein equations must
break down (inside black holes for example), and quantum field theory in
flat spacetime produces infinite amplitudes which, in the view of many
workers, are only partly explained away by renormalization.

Taken together, all these difficulties and incompatibilities have
suggested to many people that either Relativity theory or Quantum theory
or both will have to be fundamentally modified before a successful union
of the two will be achieved.

\bigskip
\noindent{\bf The causal set idea}
\nobreak

At present my main hopes for quantum gravity center on an idea (the
causal set idea) which by now has been around for a while, even if most
people haven't taken it too seriously. I imagine that one reason for
this neglect is that it is very hard to come up with plausible ``laws of
motion''  for causal sets. Conversely, one of the things that encouraged
me to begin to champion causal sets more enthusiastically was that I did
finally get a glimpse of a possible dynamics for them. Equally important
however, was the influence of M. Taketani's writings, which convinced me
that there is nothing wrong with taking a long time to understand a
structure ``kinematically'' before you have a real handle on its
dynamics.  In fact I think that Taketani's recognition of the importance
of what he calls the ``substantial'' stage in the development of
scientific understanding, allows him to put forward an analysis [2] of theory
construction which is ``non-trivial'' in a way that other analyses I have
seen are not. I might even have devoted my talk to an exposition of his
ideas, had I not been asked to speak on something relating directly to
quantum gravity. Anyhow, let me return to the topic that I am discussing,
and tell you in the first  place what the causal set concept actually is.

The idea [3] is that in the ``deep quantum regime'' of very small distances,
gravity is no longer described by a tensor field  living on a continuous
spacetime manifold (the metric field). Rather, the notions of length and
time disappear as fundamental concepts, and the manifold itself
dissolves into a discrete collection of elements related  to each other
only by a microscopic ordering that corresponds to the macroscopic
notion of before and after. Because of this correspondence the order may
be called `causal', and  the structure it describes  a `causal set'. It
is a ``discrete manifold'' (to use Riemann's term), and its defining
order carries in particular all the information showing up at larger
scales as the  geometry of continuous spacetime: the topology, the
differentiable structure,\footnote{*}{
A differentiable structure on a manifold is a technical notion of
``smoothness'';  without one, a manifold can not support a metric field
(or any other tensor field).}
and the metric.

Mathematically a causal set may be defined as a locally finite partially
ordered set, or in other words a set $C$ provided with a ``precedence''
relation, $\prec$,  subject to the following axioms:\smallskip{\narrower

  \item{(1)} if $x \prec y$ and $y \prec z$
	     then $x \prec z$  (transitivity);
  \item{(2)} if $x \prec y$ and $y \prec x$
	     then $x = y$  (non-circularity);
  \item{(3)} for any pair of fixed elements $x$ and $z$ of $C$,
	     the set $\{y |  \, x\prec y\prec z\}$ of elements lying
	     between $x$ and $z$ is finite;
  \item{(4)} $x \prec x$ for any element $x$ of $C$ (reflexivity).
\smallskip}

\noindent Of these axioms the first and second say that $\prec$ is a partial
ordering, the third expresses local finiteness, and the fourth is a standard
convention made for convenience. Instead of saying that $\prec$ is a partial
ordering, one also says that $C$ is a ``partially ordered set'', or
``poset'', for short.

In the figure I have shown three examples of rudimentary causal sets,
represented graphically in a way suggested by the spacetime diagrams of
Relativity theory. In these ``Hasse diagrams'' each ``vertex''
represents an element of the causal set, and each rising ``edge''
represents a relation.  For clarity, not all of the relations are shown
explicitly, but only those not implied via transitivity (axiom $1$) by
other relations. Thus, for example, the lowest element precedes the
highest element in the second poset even though no direct line is shown
joining them.

Of course,  a causal set underlying even a very small portion of
spacetime would be immeasurably larger than those shown, but the third
picture is meant to give some flavor, at least,  of how a realistic
causal set might look.  In contrast, the first and second posets (as
well as being too small) are probably too regular to be realistic, but
they do give some idea of how dimensional information can be present in
a causal order. The second is clearly laid out like a two-dimensional
checkerboard, and it can in fact be embedded as a subset of
two-dimensional Minkowski space.  The first has dimension three in a
certain sense, since it can be embedded in a flat spacetime only if the
latter has a dimension of three or higher.

One thing that the pictures do not show, is that macroscopic spacetime
volume is supposed to be a measure of the number of elements in the
corresponding region of the causal set.  This is a crucial part of the
physical interpretation, and a relationship that makes sense only
because of the intrinsic discreteness of the causal set.

There is much more to be said on how a causal set can manage to
determine an approximate spacetime metric, and also why a theory based
on causal sets can be expected to bypass some of the difficulties of
quantum gravity that I referred to earlier. [4] Since this is a meeting on
theory {\it construction}, however, I will not further describe or argue for
the causal set idea as such.  Instead I will try to reconstruct for
you the chain of thought which, in my case, led from certain general
expectations and desires to the particular proposal for quantum gravity
that I have just sketched. Unfortunately some of this account will be
rather technical, but I hope the general development will be clear, even
if the meaning of certain concepts and issues remains partly obscure.

\bigskip
\noindent{\bf Initial expectations}
\nobreak

The ideas from which I started were, I think, discreteness (or
``finitarity''), operationality, and a desire to negate the manifold as
the substratum of spacetime physics.

That spacetime might ultimately be discrete rather than continuous is an
idea that goes back at least to the time of Zeno.  In the last century
it was clearly enunciated by Riemann and Boltzmann [5], and it has plainly
been ``in the air'' for the last several years.  One big reason for its
recent currency is certainly the problem of infinite amplitudes in
Quantum Field Theory that I referred to earlier, and to a lesser extent
the problem in General Relativity of singularities at which the
spacetime curvature becomes infinite.  (And here I would add an
infinity which, I think, is unduly overlooked: the infinite {\it
entropy} that a black hole horizon would possess if arbitrarily fine
variations in its shape, or arbitrarily fine  fluctuations of
matter fields
in its neighborhood, were to contribute.) These
``ultraviolet'' infinities arise at infinitely short distances, and
consequently would be immediately converted to finite quantities if
there were some lower limit to the physical distances that actually
exist.

Such a potential resolution of the problem has become much more real for
physicists with the advent of so-called ``lattice gauge theories''[6],
which allow actual computations to be made on the basis of discrete
(albeit artificially constructed) spacetimes.  In fact, these discrete
spacetimes are just transpositions to four dimensions of the atomic
lattices that ordinary solids form. And conversely, when people adapted
field-theoretical methods to the understanding of
ordinary solid matter,
they obtained quantum field theories
with divergences which are manifestly no more than an artifact
of the continuum approximation being employed.  In this case, the
``cutoff'' that removes the infinities has a physical meaning which is
transparent and uncontroversial.

Thus, the atomic structure of matter  has suggested to physicists a like
character for spacetime. In a similar way, the historically unexpected
discreteness (of energy, volume in phase space, angular momentum,
$\ldots$) from which quantum mechanics gets its name also has intimated
that an underlying granularity of apparently continuous quantities is a
universal feature of nature. And finally there are digital computers.
Without them, the lattice gauge computations I just mentioned could
never have been done. But beyond that, their broader influence on
scientific culture clearly reinforces a belief in the ultimately
``finitary'' nature of the microscopic world.

The prestige of ``operationality'' as a guiding principle is another
fact of scientific culture whose roots are probably too deep to be fully
unearthed. Being a scientific form of positivism,  its presuppositions
might have been merely transferred to physicists from bourgeois
philosophy, where I think a positivistic approach has tended to dominate
in this century. Within science itself, the strongest arguments have
been based on the fact that $-$ for whatever reasons $-$  the unfolding
of the quantum and relativity revolutions of this century has commonly
been (mis)represented as a triumph of the operationalist method.

As applied to gravity, operationalism would require that the fundamental
variables be things with ``direct experimental meaning'', like the
components of the metric tensor. And it would tend to construe this
tensor as merely a {\it summary} of the behavior of idealized clocks and
measuring rods, rather than as an independent substance, whose {\it
interaction with} our instruments gives rise to clock-readings, etc.

Finally, there is the desire to transcend the manifold concept, which
has also been felt by many people, particularly those with a strong
interest in gravity.  Of course this desire is connected with the urge
toward finitarity, but it also has independent roots, which unfortunately
seem to be more technical in nature than the issues we have
just been considering. For me, I think the strongest reason
for dissatisfaction with the manifold concept
had to do
with quantum fluctuations in the microscopic topology. The occurrence of
such tiny fluctuations is strongly suggested by the form of the
Einstein-Hilbert lagrangian, but the resulting picture of ever-changing
topological complexity on infinitely small scales (the so-called `foam')
is not one that can be painted in manifold colors.  Indeed I did not
even see how, within the manifold framework, you could express
consistently the notion that topological fluctuations of {\it finite}
complexity can ``average out'' to produce an uncomplicated and smooth
structure on larger scales.  The replacing of a manifold by something
more fundamental, I felt,  might allow such a picture to make technical
sense. It might also provide a route to answering a related question
that many people have hoped quantum gravity would be able to address,
namely the old question of why there are precisely three spatial
dimensions and one temporal dimension, rather than some other number of
each. Such a possibility looked particularly attractive in light of the
revived interest in Kaluza-Klein theories, which posit that spacetime
at sufficiently small distances actually does have a different
dimensionality than that of our daily experience.\footnote{*}{
Since the time I have been talking about, my study of so-called
topological geons has convinced me even more strongly that
topology-change is an unavoidable feature of quantum gravity. Also
further evidence has accumulated that this phenomenon at least
stretches, and probably bursts through, the manifold framework. However
what I have described here is meant only as a static sketch of my
thinking at the time when the train of thought leading toward causal
sets got underway. For clarity, I have tried to exclude from my account
any supporting considerations that arose later.}

\bigskip
\noindent{\bf Simplicial gravity}
\nobreak

Led by the expectations and prejudices I have just described, I looked
for some finitary model of gravity with an operational flavor.  The only
one I found $-$ indeed the only discrete model I found at all $-$ was one in
which spacetime is represented as a simplicial complex.[7] In this
so-called ``Regge Calculus'',  which I encountered while a graduate
student, curved spacetime is replaced by an assemblage of flat
simplicial blocks, a simplex being the higher-dimensional analog of a
triangle or tetrahedron. Such an assemblage is finitary in the sense
that its geometry is fully determined by giving only a discrete list of
real numbers: one length per simplex edge. Indeed Regge calculus is
nothing but what engineers would call a ``finite element'' description
of spacetime. The flat simplexes approximate a curving manifold in just
the same way that a geodesic dome approximates the surface of a sphere.
(A finitary purist would complain that even a single real number
already contains an infinite amount of information, but in any case the
structure is discrete in the sense that there are only a finite number
of simplexes in any bounded region of the complex.)

A description in terms of simplicial complexes also has an operational
flavor.  Imagine that what we call a spacetime point is merely an ideal
limit of finer and finer experimental ``determinations of location''
(measurements). Since the actual measurements are imperfect, they will
determine not individual points but ``fuzzy'' ones corresponding roughly
to the topological concept of an open set.  Now if you cover a manifold
by a finite number of open sets, and if you keep track of which of these
sets (or determinations) overlap with each other, then you get what is
called the `nerve' of the covering, and this nerve is a simplicial
complex!  Thus one could view the use of simplicial complexes in Regge
calculus as a kind of formalization of ``what we actually do to produce
spacetime by our measurements''.

At that time, however, my adherence to the discrete camp was not
complete. I still believed in a {\it potential} continuum existing as an ideal
limit of the actual discrete, or more specifically as a limit of finer
and finer position determinations. There would thus be no bound in
principle to how precisely we could measure spacetime location, and
therefore no unbreechable minimum length in nature. Accordingly, I
thought of spacetime not as a single simplicial complex, but rather as a
sequence of them, each refining the previous one, with the whole
sequence converging in the limit to some topological space that need not
be just a manifold. This framework (described almost in these terms in
an old Dover reprint [8] by Pavel Alexandrov) promised, with its built-in
possibility of different simplicial complexes on different scales, to
give a precise meaning to the intuitive picture of topological
fluctuations that I referred to before.

But this promise was one I was never able to redeem. The simplicial
complex of Regge calculus seemed in the end to be a useful tool for
approximating the continuum theory, but not, after all, a finitary
structure that could serve as the physical underpinning of the
continuum. In Regge calculus the dynamics (or ``equation of motion'') is
given by varying the discrete analog of a lagrangian, but this analog
ceased to be meaningful as soon as you took the complex to be
more-than-slightly more general than a manifold. For this reason also,
the dimension had to be chosen in advance, and therefore seemed to be no
more explicable in the simplicial
framework than in ordinary General Relativity.
In addition there seemed to be no natural way to include Fermi fields in
the framework, although gauge fields did find their natural place.

However all these difficulties were ones which you could imagine
removing with greater ingenuity. The failing that carried the greatest
weight in my mind was actually a technical problem.  It turned out that
the successive complexes did not really converge to their putative limit
as the determinations defining them became finer and finer! To say
precisely what this means would be too much of a mathematical aside, but
the basic problem was this.  You could start with a given continuous
space  (say a manifold) and cover it by a finite collection of open
sets, each representing a fuzzy point, perhaps. By adding finer and
finer open sets, you got a sequence of simplicial complexes which did
indeed have a well-defined limit  (the so-called ``inverse limit''), but
the simplexes of the complex were not all getting small when regarded as
subsets of this limiting space. Thus there was convergence in a certain
mathematical sense, but not in the physical sense that successive
approximations were corresponding to successively smaller scales of
physical size.

\bigskip
\noindent{\bf The finitary topological space}
\nobreak

The next step beyond the continuum was to discard the simplicial complex
as the basic structure, and try instead the finite topological space. In
grappling with the limit problem I just told you about, I had noticed
that the nerve of a finite open covering does not actually encode all the
information about how the sets of the covering overlap each other; it
only keeps track of their mutual intersections.  If you do keep
all the overlap information, then you end up not with a simplicial
complex, but with a different mathematical structure: a
finite topological space.\footnote{*}{
Actually, a strictly finite covering is appropriate only for a bounded
region of spacetime. In the more general case of a locally finite
covering, as would be needed for an infinitely extended region of
spacetime, the structure you get is what might be  called a ``finit{\it
ary''} topological space, one fulfilling a certain natural condition of
{\it local} finiteness.}
Like the simplicial complex, this structure also carries information of
a topological nature. (In fact it {\it is} literally a topology, as its
name says.)  But  unlike before, the sequence of finite topological
spaces corresponding to a sequence of finite open coverings {\it does}
converge properly to the continuous space being approximated. This
seemed to open the way for a true negation of the spacetime manifold,
something whose discrete/combinatorial character was more thoroughgoing
than that of the Regge calculus had proved to be. It thus seemed
possible that, of all the structures defining a smooth manifold, it
would turn out to be the topology itself that bears the greatest
structural similarity to the underlying discrete reality. [9]

As the correspondence between open coverings and finite topologies was
becoming clear to me, I also realized that a finite topological space
has a very different, yet entirely equivalent, description as a partial
ordering. This intriguing correspondence between topology and order
struck me as deep in itself; but it also resonated in my mind with a
tradition in physics and philosophy of wanting to base the analysis of
spacetime structure on the properties of a quite distinct
order-relation $-$ the
so-called causal order of ``past'' and ``future'', which in standard
General Relativity tells you which events are able to signal to (or more
generally to influence) which other events.  Still, the order inhering
in the finite topological space seemed to be very different from the
order defining past and future. It had  only a topological meaning but
not (directly anyway) a causal one.

In fact the big problem with the finite topological space was that it
seemed
to lack the kind of
information which would allow it
to give rise to the continuum
in all its aspects,
not just in its topological aspect, but
with its metrical (and therefore its causal)
properties as well. Could it be, then, that everything
is ultimately topological $-$ that even the notions of length and time
emerge somehow from more fundamental relations of adjacency and
convergence? To address this question I tried (maybe not very hard) to
make a theory of dynamical topology alone (i.e. I tried to find a
quantum law of motion for the finite topological space), but I got
nowhere.  On the other hand, the only way I could see to put metrical
information back in explicitly, was to use a certain correspondence that
exists between finite topological spaces and simplicial complexes, and
then appeal to Regge calculus.  But this would put us back where we
started, not having gotten essentially beyond the manifold concept.

\bigskip
\noindent{\bf The causal set}
\nobreak

The way out of the impasse involved a conceptual jump in which the formal
mathematical structure remained constant, but its physical interpretation
changed from a topological to a causal one.  Although, unfortunately, I
can no longer recall the inner development of this jump in any detail, it
is easy to see it in retrospect as a natural step.

I wrote above that the mathematical structure ``finite topological space''
is equivalent\footnote{*}%
{ Strictly speaking, this equivalence presupposes that distinct elements
of the topological space possess distinct neighborhood systems (that the
topology is what is called $T_0$).}
to the structure ``partially ordered set''; and  in fact I
normally thought about finite topologies in the latter language, since it
seemed to provide the more convenient representation in most cases.  I was
thus already thinking of the fundamental discrete structure as an order
(poset), but an order with a topological meaning.  The essential
realization then was that, although order interpreted as topology seemed
to lack the metric information needed to describe gravity, the very same
order reinterpreted as a causal relationship, did possess metric
information in a quite straightforward sense.

Or rather, it possessed the necessary information if you abandoned
operationalism and took the causal set to be a real substratum, existing
independently of any experimental activity on our part.  This
meant accepting an actual
minimum length in nature, and it made possible the key hypothesis that I
referred to earlier of a direct proportionality between number and
volume.  By itself, the choice of a causal ordering as basic could have
been more than justified in operational terms, but there is nothing in
what we do when we measure spacetime volume that phenomenologically has
the character of counting.  To believe such a relationship, you have to
accept that the elements of the causal set are real, and that volume
measurements ``count'' them in much the same way that weighing a copper
ingot ``counts'' the number of atoms it comprises.

[Indeed, weighing is not precisely the same as counting atoms; and  I
would not view as {\it exactly} true either
half of the compound hypothesis
that microscopic number shows up macroscopically as spacetime volume,
and microscopic order shows up macroscopically as causality. Rather I
like to think of these basic assumptions as analogous to the
hypothesis in classical General Relativity that bodies move along
spacetime geodesics.  Viewed thusly, they would belong to what Taketani
would call the ``substantialistic'' stage of understanding of quantum
gravity, being essentially approximate relationships, which will be
corrected and more fully understood only in the higher theoretical stage
when an exact dynamics is available.]

The result of these changes was that now you no longer needed to add
anything to the combinatorial data, in order to recover the metrical
aspects of the continuum. Everything necessary for gravity was already
present in the unadorned causal set, whose discreteness, moreover, was
now intrinsic to the physical interpretation (thereby realizing
Riemann's claim, that for a discrete manifold, metrical properties do
not need to be added in by hand.) Potentially unified now, in terms of a
single notion of microscopic order, were all the basic structures going
into the General Relativistic conception of the continuum $-$ its topology,
its differentiable structure,
its metric and its causal structure.
In addition, the Lorentzian signature of the metric tensor $-$ in other
words, the fact that precisely one of the dimensions is timelike with
all the other ones being spacelike $-$ became inevitable, whereas in
itself it appears mathematically unnatural and inconvenient.

As I just said, I am not sure exactly how these changes in my thinking
took place, but one stimulus for the transition from topological order
to causal order may have been my going to Chicago, where David Malament
had just emphasized how much information the causal ordering actually
contains in the continuum case. [10] This also may have highlighted for me
what the continuum order {\it fails} to contain $-$ namely information
on spacetime volume $-$ and may thereby have prepared a sudden
realization that my discrete order could make up for this lack if
reinterpreted in a causal way.  My retreat from operationalism, on the
other hand, was definitely not sudden. It  was part of a much slower
evolution with causes partly inside and partly outside of physics
proper.

The reasons for accepting the causal set as the right structure sound
convincing to me now, but for a long time  I remained in some doubt.  In
fact it took me several years to definitively give up the idea of
order-as-topology and adopt the causal set alternative as the
one I had been searching for.  Whether the resulting hypothesis is true,
can of course be decided only after a lot of further work.

\bigskip
\noindent{\bf Other threads: fermions, geons, the
sum-over-histories, $\ldots$}
\nobreak

In essence this is the end of the story, but there remains at least the
question that Dirac was said to have always asked when he woke up at the
end of a seminar: ``but what about the muon'' $-$ or in this case, what
about fermions in general?  Earlier I mentioned that one of the
difficulties of the Regge calculus was that it did not appear to be able
to accommodate fermi fields in a natural way, whereas fermions certainly
exist in the real world. As far as I know, the situation is no better
with causal sets, but in the meantime I have found out that I was wrong
in thinking that fermions must occur at the fundamental level of any
theory in which they occur at all. In principle they can all emerge at a
higher level, as composite particles, or as objects derived in some less
obvious manner from the fundamental structures. In fact you don't even
need to go beyond pure gravity to get fermions since they can occur as
topological excitations (``geons'') in four dimensions [11], or more
exotically, on the basis of the higher dimensional manifolds of
Kaluza-Klein theory (Kaluza-Klein geons). This was a parallel
development, and to some extent a hidden thread in the story I have been
recounting.

Other hidden threads in the story concern the so-called
sum-over-histories (or ``path integral'') interpretation of quantum
mechanics in general [12], the conflict in my thinking between a more
``algebraic/logical'' and a more ``geometrical/set theoretic'' approach
to the quantum gravity problem, and the question of whether some
conceptual descendant of the pre-Relativistic notion of time will
continue to play a role in the dynamical ``law of motion'' of quantum
gravity.  These threads intertwine with each other and with the fermion
issue as well, but there is a limit to how tangled a history you can
tell, or even begin to reconstruct in your own\break
mind $\ldots\ldots$

\medskip
\noindent
This research was partly supported by NSF grant PHY 8700651.

\parskip=10pt
\parindent=0pt
\vskip 0.5truein
\centerline {\bf References}
\nobreak
\medskip\noindent

[1]  Finkelstein, D., ``The space-time code'', {\it Phys. Rev.} {\bf
184}, 1261 (1969);
Myrheim, J., ``Statistical geometry,'' CERN preprint
TH-2538 (1978);
't Hooft, G., ``Quantum gravity: a fundamental problem
and some radical ideas'', in {\it Recent Developments in Gravitation }
(Proceedings of the 1978 Cargese Summer Institute) edited by M. Levy and
S. Deser (Plenum, 1979).

[2] see the articles by M. Taketani and S. Sakata in
{\it Suppl. Prog. Theor. Phys.} {\bf 50} (1971), especially the article
by Taketani on Newtonian mechanics.

[3]  Bombelli, L., Lee J., Meyer, D. and Sorkin, R.D., ``Spacetime as a
causal set'', {\it Phys. Rev. Lett.} {\bf 59}, 521 (1987).

[4]  Bombelli, L., {\it Space-time as a Causal Set}, Ph.D. thesis, Syracuse
University (1987);
Meyer, D.A., {\it The Dimension of Causal Sets}. Ph.D. thesis, M.I.T.
(1988);
Moore, C., ``Comment on `Space-time as a causal set','' {\it Phys. Rev.
Lett.} {\bf 60}, 655 (1988); and the reply in {\it Phys. Rev. Lett.} {\bf
60}, 656 (1988).

[5] Riemann, G.F.B., {\it \" Uber die Hypothesen, welche der Geometrie
zugrunde liegen,} edited by H. Weyl (Berlin, Springer-Verlag, 1919); and
Boltzmann, L., {\it Theoretical Physics and Philosophical Problems
(selected writings)}, edited by B. McGuinness (D. Reidel, 1974).

[6] {\it Advances in Lattice Gauge Theory}, edited by D.W. Duke and
I.F. Owens (Singapore, World Scientific, 1985).

[7] Regge, T., ``General Relativity without Coordinates'',  {\it Nuovo
Cimento} {\bf 19}, 558 (1961);
the basic idea is also explained in:
Sorkin, R.D., {\it Development of Simplicial Methods for the Metrical and
Electromagnetic Fields}, Ph.D. Thesis, California Institute of Technology
(1974) (available from University Microfilms, Ann Arbor, Michigan); and in
Sorkin, R.D., ``The Time-evolution Problem in Regge Calculus'', {\it Phys.
Rev.} {\bf D12}, 385-396 (1974); and {\bf D23}, 565 (1981).

[8] Alexsandroff, P., {\it Elementary Concepts of Topology}, (Dover,
1961).

[9] for a recent exposition of some of these ideas see: Sorkin, R.D., ``A
Finitary Substitute for Continuous Topology?'', {\it Int. J. Th. Phys.},
{\bf 30}, 923-947 (1991)

[10]  Malament, D., ``The class of continuous timelike curves determines the
topology of space-time'', {\it J. Math. Phys} {\bf 18}, 1399 (1977).

[11] Friedman, J.L. and R.D. Sorkin, ``Spin-$1/2$ from Gravity'',
{\it Phys. Rev. Lett.} {\bf 44}, 1100-1103 (1980); and {\bf 45}, 148 (1980)

[12] Feynman, R.P., ``Space-time Approach to Non-relativistic Quantum
Mechanics'', {\it Rev. Mod. Phys.} {\bf 20}, 267 (1948);
for an explicit illustration that no ``wave-function collapse'' is needed
to account for the so-called Einstein-Podolsky-Rosen correlations in this
framework, see Sukanya Sinha and R.D. Sorkin, ``A Sum Over Histories
Account Of An EPR(B) Experiment'', {\it Found. of Phys. Lett.},
{\bf 4}, 303-335 (1991)

\eject
\bigskip
FOR COMMENTS:

\bigskip
algebraic/logical refers to the old ``quantize metric'' work, in which
$g_{\mu\nu}$ and $x^\mu$ would be q-numbers; there would be
``chaotic'' variables, etc.

\end